\documentclass[%
reprint,
showpacs,preprintnumbers,
amsmath,amssymb,
aps,
pre,
superscriptaddress
]{revtex4-1}

\usepackage{soul}
\usepackage{graphicx}
\usepackage{dcolumn}
\usepackage{bm}
\usepackage{braket}
\usepackage{color}
\usepackage{caption}
\usepackage{graphicx}
\usepackage{ragged2e}
\usepackage{natbib}
\usepackage{hyperref}
\usepackage{subcaption}
\usepackage{url}
\usepackage[utf8]{inputenc}
\usepackage{soul}
\usepackage{amsmath}
\usepackage{comment}
\usepackage{color}
\usepackage{subcaption}
\DeclareCaptionFormat{myformat}{#1#2(Color online) #3}
\DeclareCaptionJustification{justified}{\justifying} 
{\captionsetup{format=myformat,labelsep=period,justification=justified, font=small}
\figure
}%
{\endfigure}%

\begin{document}
\preprint{APS/123-QED}
\title{Solid state emulation of the photosynthetic reaction center}
\author{Vishvendra Singh Poonia}
\email{vspfec@iitr.ac.in}
\affiliation{Department of Electronics and Communication Engineering, Indian Institute of Technology Roorkee, Haridwar -- 247667, India}


\date{\today}%

\begin{abstract}
The photosynthetic reaction center of plants and bacteria is an extremely efficient energy to charge conversion device. Solar photons create excitons in the pigment molecules. These excitons are then transferred to the reaction center where charge separation takes place. These processes - excitonic generation and subsequent charge separation are extremely efficient with almost unity efficiency. Taking pointers from this biophysical process, we propose a GaN quantum dot based solid state energy to charge conversion device idea that emulates the photosynthetic reaction center. This further suggests that highly efficient quantum biological processes can give important pointers for developing energy harvesting quantum technologies.

\end{abstract}

\pacs{Valid PACS appear here}

\maketitle
\section{Introduction}
\label{Intro}
Photosynthesis is the process by which various organisms such as plants, algae, and some species of bacteria convert solar energy into the usable chemical energy. It, directly or indirectly, powers almost all living beings on planet Earth. In this process, carbohydrates are synthesized from carbon dioxide and water using sunlight. The process of photosynthesis is broadly constituted by the following steps:
\begin{itemize}
 \item 
Solar photons strike the pigment molecules and generate excitons (bound state of electron and hole).
 \item Subsequently, the antenna complex network transfers these excitons to the reaction center where charge separation takes place.
 \item Charge separation is enabled by a series of electron transfer reactions that eventually help in carbohydrate synthesis.
\end{itemize}

Exact details of the molecular reactions involved in photosynthesis have been furnished in great detail in Ref.~\citep{blankenship2013molecular}. Depending on whether the photosynthetic chemical reactions release oxygen as byproduct or not, they are classified as oxygenic or anoxygenic photosynthetic reactions. A wide variety of organisms are involved in photosynthesis. Apart from most of the plants and algae, some bacterial species and protists also employ photosynthesis for energy generation~\citep{blankenship2006anoxygenic,yoon2004molecular}. The mechanism of bacterial photosynthesis is among the most well-studied photosynthetic mechanisms owing to relatively simpler photosynthetic apparatus of bacteria. 



\section{Efficiency of photosynthesis}
Of all the photons that are incident on the photosynthetic apparatus, only a very small fraction of them is eventually converted and stored in the form of chemical energy. Following are the dissipation mechanisms in the plant photosynthesis~\citep{Hall1999Photosynthesis}:
\begin{itemize}
\item Plants absorb light in the wavelength range of 400 - 700 nm. Around 47\% of the incident light falls out of this range and hence not utilized in photosynthesis.
\item Further, 30\% of the absorbed light is lost due to photons hitting other than the chlorophyll molecules in the pigments.
\item Of all the photons transported through pigment protein complex (PPC) of the photosynthetic apparatus, 24\% of them are lost because chlorophyll molecules only absorb the energy equal to the energy of a photon of 700 nm wavelength from each photon. Therefore, several low energy photons are lost in the process.
\item 28\% of the energy is further lost in the process of converting it into d-glucose.
\item 35\% - 45\% of the produced glucose is consumed by the leaves in the process of respiration.
\end{itemize}
Hence the overall efficiency of the photosynthesis turns out to be 3 -- 6\%. Despite the low overall efficiency of the photosynthesis, the transport of excitonic energy through the antenna complex and charge separation at the reaction center are highly efficient mechanisms with their efficiency approaching almost unity~\citep{blankenship2013molecular}.

\section{The photosynthetic apparatus}
\begin{figure}[t]
\centering
\includegraphics[width=8cm,keepaspectratio]{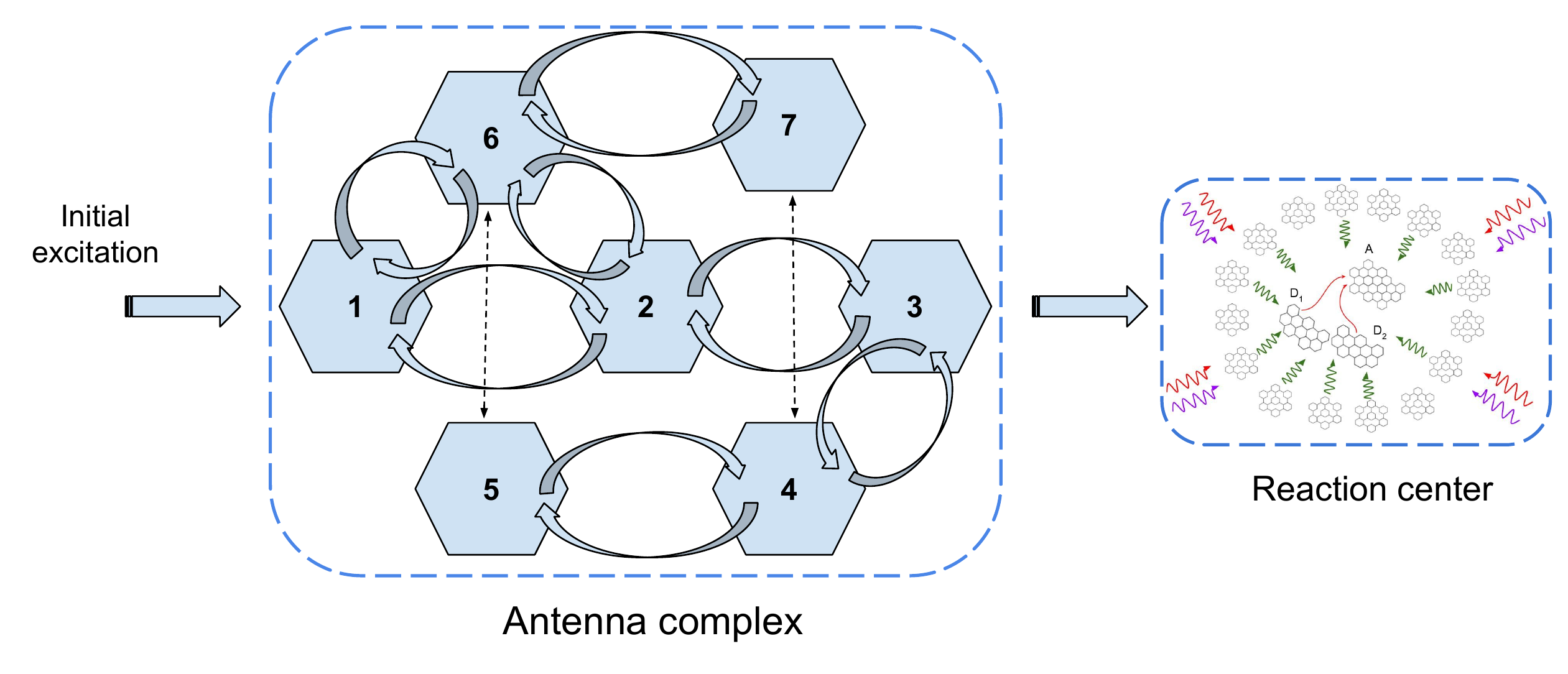}
\caption[An illustration of the photosynthetic apparatus]{A general illustration of the photosynthetic apparatus. It mainly contains two components: a) the antenna complex and b) reaction center. Sunlight is absorbed by pigments and transported to the reaction center via antenna complex in the form of excitons. The process of charge separation takes place at the reaction center.} 
\label{PSApparatus}
\end{figure}

In this work, we are only interested in the light dependent part of photosynthetic reaction. Two main components of the photosynthetic apparatus are involved in this part:
\begin{enumerate}
\item \textbf{Antenna complex: }
Incoming solar photons create excitons in the pigment molecules. The antenna complex of the photosynthetic apparatus is responsible for transporting these excitons to the reaction center in highly efficient manner.
\item \textbf{Reaction center: }The reaction center captures the excitonic flux from the antenna complex. It is responsible for charge separation and supplying electrons for further chemical reactions.
\end{enumerate}
Fig.~\ref{PSApparatus} illustrates a typical photosynthetic apparatus. The process of photosynthesis starts with the generation of excitons due to incoming photons in the pigment molecules. The antenna complex consists of a network of chlorophyll molecules which is responsible for the transport of excitons from pigment molecules to the reaction center. In the reaction center, the excitonic energy is converted into more stable form of electrochemical energy. The exact structure of the antenna complex and the reaction center of an organism depends heavily on its habitat and physiological conditions it encounters there. A wide variation is observed in the structure of the photosynthetic apparatus of different organisms e.g. purple sulphur bacteria has ring-like structure for light harvesting antenna complex~\citep{lambert2013quantum,cogdell2006architecture} whereas the antenna complex of green plants and cyanobacteria consists of randomly arranged photosystems with chlorophyll molecules~\citep{lambert2013quantum}. 

In our work, we only deal with the charge separation mechanism in the reaction center which is elaborated in the next section.

\section{The reaction center and its quantum heat engine model}
\begin{figure}[t]
\centering
\includegraphics[width=7cm,keepaspectratio]{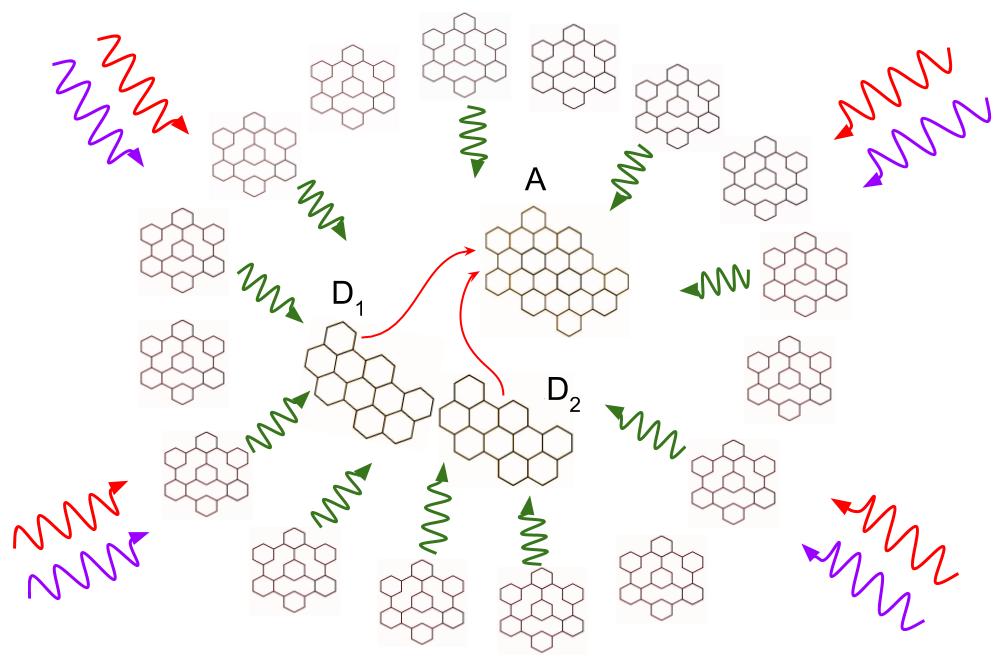}
\caption[Charge separation mechanism in photosynthetic reaction center]{Mechanism of charge separation in reaction center is shown in this figure. The reaction center consists of a pair of donor molecules that receive excitons from the antenna complex. The donors transfer the electron to a molecule known as reaction center. These electrons are further used in chemical reactions where the synthesis of carbohydrates takes place.} 
\label{PSReactionCenter}
\end{figure}

After the sunlight has been absorbed by pigments on the antenna complex, the generated excitons are transported to the reaction center where the charge separation takes place, as illustrated in Fig.~\ref{PSReactionCenter}.

At the core of reaction center complex, there lies two dominant charge separation pathways in both bacterial and plant photosynthetic reaction centers~\citep{deisenhofer1985structure,romero2010two}. This implies that there are at least two donor molecules that receive excitons from the antenna complex and donate the electrons to an acceptor molecule~\citep{dorfman2013photosynthetic}. The excitonic dissociation takes place at this part of the reaction center. The schematic of a typical photosynthetic reaction center is given in Fig.~\ref{PSReactionCenter}. A few recent studies have suggested that coherence between the charge separation pathways is responsible for high efficiency of the charge separation~\citep{dorfman2013photosynthetic,creatore2013efficient}.

\begin{figure}[t]
\centering
\includegraphics[width=7cm,keepaspectratio]{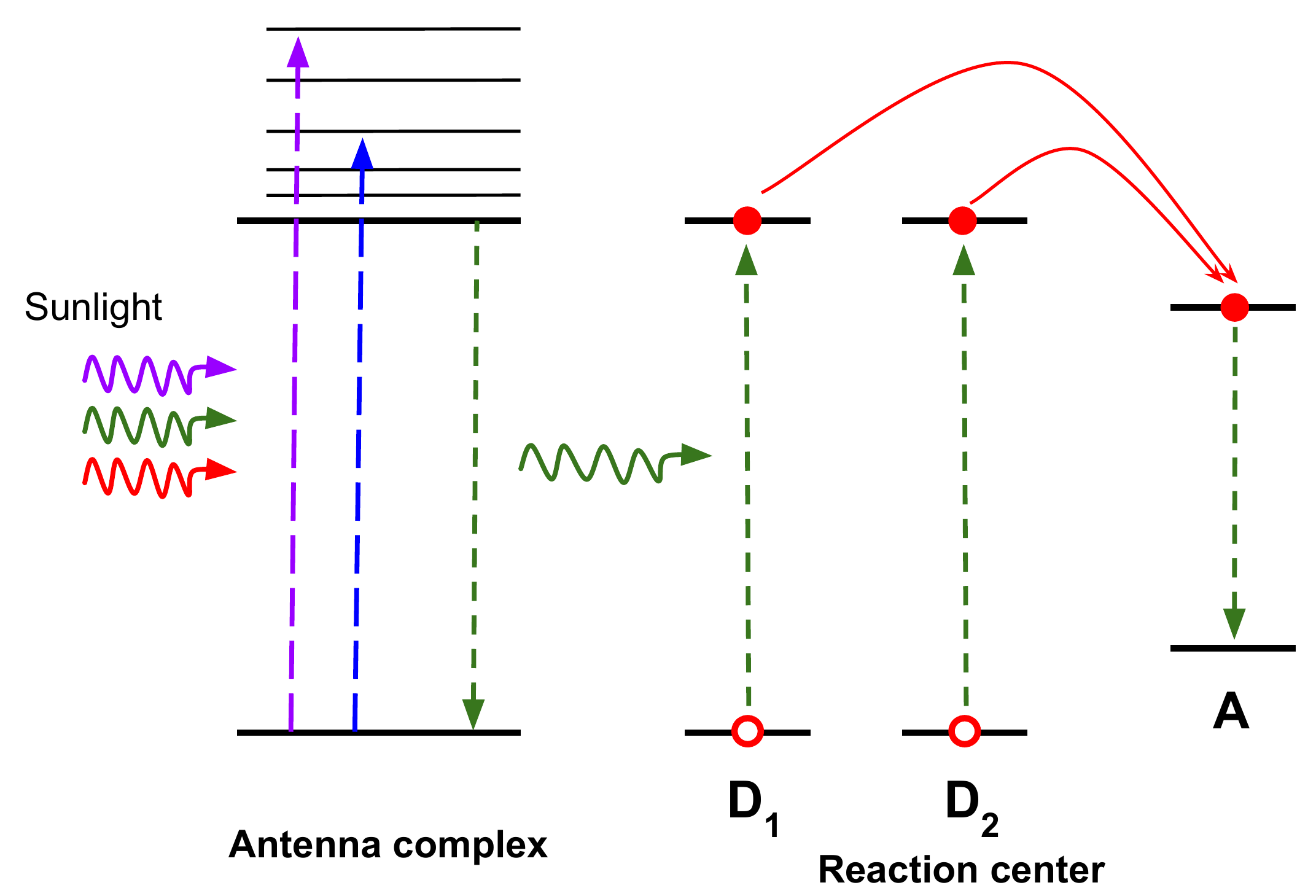}
\caption[Dynamics of excitonic transport and charge separation in photosynthetic apparatus]{Dynamics of excitonic transport and charge separation in photosynthetic apparatus is shown. Broadband sunlight is absorbed by the pigments. Subsequently, the excitonic energy is transported via the antenna complex and narroband excitation is transferred to the `special' pair of reaction center. Overall, the photosynthetic reaction center converts the stream of photons (sunlight) into electron flux and can thus be visualized as a heat engine.} 
\label{RC_QHEModel}
\end{figure}

The mechanism of the reaction center can be visualized as a quantum heat engine operating between high temperature solar radiation and low temperature protein phononic bath~\citep{dorfman2013photosynthetic}. It transforms the energy of the solar radiation into electron flux~\citep{dorfman2013photosynthetic}. 
The reaction center consists of a pair of donor molecules and an acceptor molecule. These molecules sit at the end of the antenna complex that transport excitons from pigments to the reaction center. Essentially, the mechanism at the reaction center can be summarized as follows:
\begin{itemize}
\item The broadband sunlight is captured by the pigments.
\item The absorbed light is then transferred to the reaction center via antenna complex in the form of narrowband excitations.
\item The pair of donor molecules in the reaction center receives these excitations and transfers electrons to an acceptor molecule sitting nearby.
\item The electrons are utilized in photosynthetic chemical reactions and we are left with the positively charged system.
\item Another electron transfer process happens from nearby environment to the photosynthetic system; thus completing  the cycle and bringing the system to its original charge neutral state~\citep{dorfman2013photosynthetic}.
\end{itemize}
This entire mechanism can be modeled in the following way~\citep{dorfman2013photosynthetic}:
\begin{itemize}
\item The initial state of the reaction center when it does not have any exciton is denoted as $a$ and the states after each of the donor molecules have received excitons are denoted as $b_1$ and $b_2$ respectively. Thus, in states $b_1$ and $b_2$, donor 1 and donor 2 molecules have bound electrons and holes.
\item After the donor molecules have transferred the electron to the acceptor molecule, the state of the reaction center is denoted as $\alpha$. Finally, $\beta$ is the state of the reaction center in which the electron from the acceptor molecule has been transferred to a ``sink" where it is utilized in photosynthetic chemical reactions.
\item The transition from the state $a$ to $b_1$ and $b_2$ is caused by the solar radiation. The transition from $b_1$ and $b_2$ to $\alpha$ happens after the electron looses excess energy in the form of a phonon. Further, the electron is transferred from state $\alpha$ to $\beta$ with a rate of $\Gamma$.
\item The final step that brings the system back to the charge neutral state is modeled as a transition from the state $\beta$ to the state $a$.
\end{itemize}
An illustration of mechanism of charge separation in the reaction center is shown in Fig.~\ref{RC_QHEModel}. The state diagram of the reaction center is shown in Fig.~\ref{RCEnergyLevels} (a).



\section{Dynamics of charge separation in reaction center -- the master equation}
In the photosynthetic reaction center, there are two charge separation pathways corresponding to the electron transfer from two donors to an acceptor. This electron transfer can either happen individually from donor molecules to the acceptor molecule or the excitons could be delocalized over the donor molecules and the charge separation might happen coherently from donors to the acceptor i.e. simultaneously from both the donors to the acceptor. First we describe the excitonic dynamics in the reaction center and then the effect of coherence on the efficiency of charge separation will be analyzed. 


\begin{figure}[t]
\centering
\includegraphics[width=8cm,keepaspectratio]{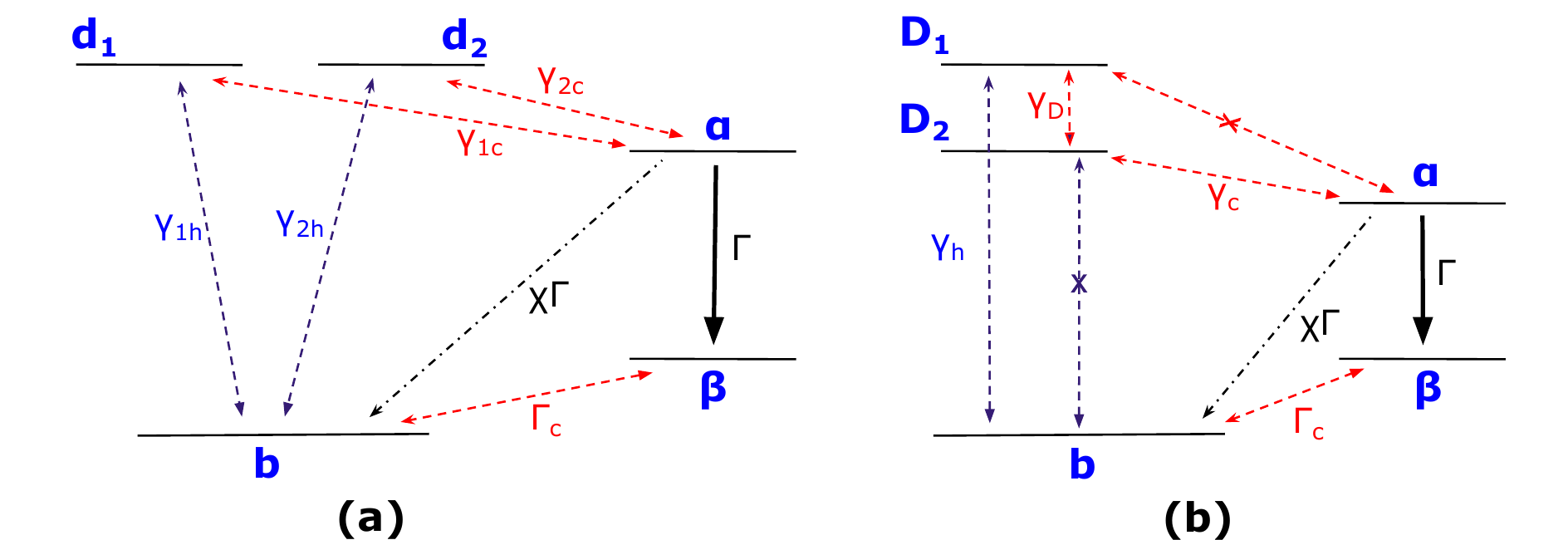}
\caption[State diagrams of the reaction center in absence and presence of dipole-dipole coupling between the donor molecules]{State diagrams of the reaction center in absence and presence of dipole-dipole coupling between the donor molecules. The coupling delocalizes the excitons over the donors. Various pathways describe the transitions among the energy levels of the center. 
}
\label{RCEnergyLevels}
\end{figure}

The donor molecules in the reaction center are very closely spaced. Therefore, the energy levels of these molecules might interfere with each other owing to the presence of excitonic dipole on them and can give rise to new energy levels after constructive and destructive interference~\citep{creatore2013efficient}. 
Alternatively, the two donor levels might undergo Fano interference induced quantum coherence owing to their coupling to the antenna complex~\citep{dorfman2013photosynthetic}. In both the cases, it has been observed that the charge separation efficiency enhances owing to these quantum coherence mechanisms. Let's take the case when donor molecules undergo excitonic coupling due to dipole-dipole interaction. In this case, the state diagram of the system looks like Fig.~\ref{RCEnergyLevels}~\citep{creatore2013efficient}. The diagram shows the reaction center energy levels without and with dipole-dipole interaction.

Due to dipole-dipole interaction between the excitonic states, the dipole moment of the new states formed after interaction is modified. This results in modified transition dipole moments of the states, with the state undergoing constructive interference has increased transition dipole moment and the state undergoing destructive interference has decreased transition dipole moment~\citep{creatore2013efficient}. This results in former state being optically much more active that the latter. In Fig.~\ref{RCEnergyLevels}, this is shown by various inter-state transitions. The figure shows that when there is no dipole-dipole interaction, there is photon mediated transition between the ground state of the system ($b$) and the excited states of both the donors ($d_1$ and $d_2$) with rates $\gamma_{1h}$ and $\gamma_{2h}$ respectively. On the other hand, after the dipole-dipole interaction between the donor pair, the state $D_2$ has very less transition dipole moment. Therefore, transition from the ground state ($b$) to this state is suppressed (shown by a crossed arrow). Now the only photonic transition happening is from the ground state ($b$) to the state of higher transition dipole moment ($D_1$). However, the charge transfer coefficient of state $D_2$ is much higher than the state $D_1$ which makes transfer of electron from state $D_1$ to the acceptor very weak~\citep{creatore2013efficient}. Therefore, a phonon mediated interaction between $D_1$ and $D_2$ brings the exciton from state $D_1$ to the state $D_2$ with the rate of $\gamma_D$. The state $D_2$ transfers the electron to the acceptor (represented by the state $\alpha$) with the rate of $\gamma_c$. The acceptor molecule transfers the electron for further chemical reactions with $\Gamma$ rate. This is the electron flux that the reaction center generates out of the incoming excitons. This whole process leaves the donors + acceptor system in a charged state. This system can pick up an electron from the surroundings and become charge neutral again, thereby going back to its ground state. This amounts to transition from state $\beta$ to the ground state of donors ($b$). The rate of this transition is denoted by $\Gamma_C$. The possibility of acceptor molecule transferring the electron back to the donors has also been accounted for with a rate of $\chi \Gamma$ where $\chi$ is a dimensionless ``loss" coefficient.

Finally, dynamics of the whole system reads as~\citep{creatore2013efficient}:
\begin{eqnarray}
\begin{aligned}
\dot{\rho}_{D_1D_1} = - \gamma_D[(1+n_D)\rho_{D_1D_1} - n_D \rho_{D_2D_2}] \\ - \gamma_h [(1+n_h)\rho_{D_1D_1} - n_h \rho_{bb}]
\label{PME1_Creatore}
\end{aligned}
\end{eqnarray}
\begin{eqnarray}
\begin{aligned}
\dot{\rho}_{D_2D_2} = \gamma_D[(1+n_D)\rho_{D_1D_1} - n_D \rho_{D_2D_2}] \\ - \gamma_c [(1+n_{2c})\rho_{D_2D_2} - n_{2c} \rho_{\alpha \alpha}]
\label{PME2_Creatore}
\end{aligned}
\end{eqnarray}

\begin{equation}
\centering
\dot{\rho}_{\alpha \alpha} = \gamma_c [(1+n_{2c})\rho_{D_2D_2} - n_{2c} \rho_{\alpha \alpha}] - (\Gamma + \chi \Gamma)\rho_{\alpha \alpha}
\label{PME3_Creatore}
\end{equation}
\begin{equation}
\centering
\dot{\rho}_{\beta \beta} = \Gamma \rho_{\alpha \alpha} - \Gamma_c[(1+N_c)\rho_{\beta \beta} - N_c \rho_{bb}] 
\label{PME4_Creatore}
\end{equation}
The population conservation dictates:
\begin{equation}
\centering
\rho_{D_1D_1} + \rho_{D_2D_2} + \rho_{\alpha \alpha} + \rho_{\beta \beta} + \rho_{bb} = 1
\label{PME5_Creatore}
\end{equation}
The thermal occupation of states is given by the Planck distribution:
\begin{equation}
n=\frac{1}{e^{\frac{\Delta E}{k_B T_a}}-1}
\label{OccupationNo}
\end{equation}
The charge separation efficiency is quantified by analyzing the current-voltage (I-V) and power-voltage (P-V) characteristics of the system. 
The current (I) in the reaction center is defined as the rate of outgoing electrons from the acceptor molecule. Mathematically, 
\begin{equation}
I = e \Gamma \rho_{\alpha \alpha}
\label{RCQHE_Current}
\end{equation}
The outgoing electrons from the reaction center are modeled as electron transfer between state ($\alpha$) and state ($\beta$). This electron transfer process is driven by the chemical potential difference between these two energy levels. The population of the states $\alpha$ and $\beta$ can be deduced from the Fermi-Dirac distribution as~\citep{scully2011quantum}:
\begin{eqnarray}
\rho_{\alpha \alpha} = \frac{1}{exp(\frac{E_{\alpha}-\mu_{\alpha}}{kT_a})+1} \\
\rho_{\beta \beta} = \frac{1}{exp(\frac{E_{\beta}-\mu_{\beta}}{kT_a})+1}
\end{eqnarray}
The voltage across states $\alpha$ and $\beta$ is given as (approximating Fermi-Dirac distribution by Maxwell-Boltzmann distribution)~\citep{scully2011quantum}:
\begin{equation}
eV = \mu_{\alpha} - \mu_{\beta} = E_{\alpha} - E_{\beta} + kT_a ln\Big(\frac{\rho_{\alpha \alpha}}{\rho_{\beta \beta}}\Big)
\label{RCQHE_Voltage}
\end{equation}
Power (P) is the multiplication of the current and voltage. $E_{\alpha}$ and $E_{\beta}$ are the energy levels of states $\alpha$ and $\beta$ respectively, as shown in Fig.~\ref{RCEnergyLevels}.
$T_{a}$ is the ambient temperature. In summary, the voltage defined by Eq.~\ref{RCQHE_Voltage} signifies the potential drop that an electron undergoes when it is transferred from state $\alpha$ to the state $\beta$~\citep{dorfman2013photosynthetic}. The current defined by Eq.~\ref{RCQHE_Current} captures rate of electron transfer from the acceptor molecule to the photosynthetic chemical reactions. 
\section{Role of coherence in charge separation mechanism in photosynthetic reaction center}
In order to observe the effect of dipole-dipole interaction and excitonic delocalization on the donors, we analyze the current-voltage (I-V) and power-voltage (P-V) characteristics of the system with and without excitonic coupling. The value of energy levels is~\citep{creatore2013efficient}: 
$E_1 - E_b = E_2 - E_b = 1.8$ eV, 
$E_1 - E_{\alpha} = E_2 - E_{\alpha} = E_{\beta} - E_{\alpha} = 0.2$ eV; the coupling is: $J_{12} = 0.015$ eV, and the rates are: $\gamma_{h} = 2\gamma_{1h} = 2\gamma_{2h} = 1.24 \times 10^{-6}$ eV, $\Gamma = 0.124$ eV, $\Gamma_c = 0.0248$ eV; the occupation numbers of ambient phonons at room temperature for various energy levels are: $n_D = 0.46, n_h = n_{1h} = n_{2h} = 60000$. \\
Without excitonic coupling, the master equations take the  form~\citep{creatore2013efficient}: \\
\begin{eqnarray}
\begin{aligned}
\dot{\rho}_{d_1d_1} = - \gamma_{1h}[(1+n_{1h})\rho_{d_1d_1} - n_{1h} \rho_{bb}] \\ - \gamma_{1c} [(1+n_{1c})\rho_{d_1d_1} - n_{1c} \rho_{\alpha \alpha}]
\label{PME1_WOExCoupling}
\end{aligned}
\end{eqnarray}
\begin{eqnarray}
\begin{aligned}
\dot{\rho}_{d_2d_2} = -\gamma_{2h}[(1+n_{2h})\rho_{d_2d_2} - n_{2h} \rho_{bb}] \\ - \gamma_{2c} [(1+n_{2c})\rho_{d_2d_2} - n_{2c} \rho_{\alpha \alpha}]
\label{PME2_WOExCoupling}
\end{aligned}
\end{eqnarray}
\begin{eqnarray}
\begin{aligned}
\dot{\rho}_{\alpha \alpha} = \gamma_{1c} [(1+n_{1c})\rho_{d_1d_1} - n_{1c} \rho_{\alpha \alpha}] \\ + \gamma_{2c} [(1+n_{2c})\rho_{d_2d_2} - n_{2c} \rho_{\alpha \alpha}] - (\Gamma + \chi \Gamma)\rho_{\alpha \alpha}
\label{PME3_WOExCoupling}
\end{aligned}
\end{eqnarray}

\begin{equation}
\centering
\dot{\rho}_{\beta \beta} = \Gamma \rho_{\alpha \alpha} - \Gamma_c[(1+N_c)\rho_{\beta \beta} - N_c \rho_{bb}] 
\label{PME4_WOExCoupling}
\end{equation}
And, the population conservation gives:
\begin{equation}
\centering
\rho_{d_1d_1} + \rho_{d_2d_2} + \rho_{\alpha \alpha} + \rho_{\beta \beta} + \rho_{bb} = 1
\label{PME5_WOExCoupling}
\end{equation}

\begin{figure}[t]
\centering
\includegraphics[width=5cm,keepaspectratio]{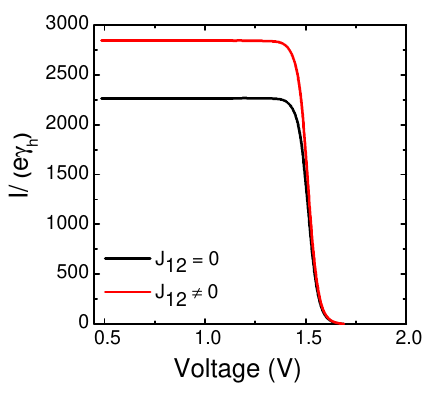}
\includegraphics[width=5cm,keepaspectratio]{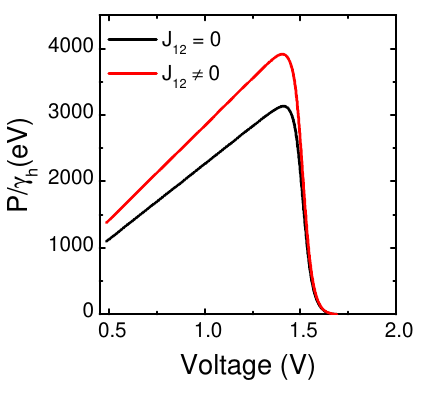}
\caption[I-V characteristics of the charge separation mechanism in reaction center with and without dipole-dipole interaction between the donor molecules]{Current-voltage (I-V) and Power-Voltage (P-V) characteristics of the reaction center with and without dipole-dipole interaction between the donor molecules. These characteristics are reproduced from reference~\citep{creatore2013efficient} and shows that excitonic delocalization due to dipole-dipole interaction might account for the high efficiency of charge separation (characterized by I-V and P-V characteristics here) in the reaction center. The figure shows an improvement of around 25\% in both the steady state current and the peak power delivered by the reaction center.}
\label{Creatore_IV_PV}
\end{figure}
The current-voltage (I-V) and power-voltage (P-V) characteristics are shown in Fig.~\ref{Creatore_IV_PV}. When the excitons are coupled via dipole-dipole interaction and are delocalized, the excitonic dynamics of the reaction center is governed by Eq.~\ref{PME1_Creatore} to Eq.~\ref{PME5_Creatore}. As is evident from Fig.~\ref{Creatore_IV_PV}, the I-V characteristics improves by around 25\% when the excitons are delocalized due to dipole-dipole interaction. Similarly, the P-V characteristics of the center also shows 25\% enhancement. This enhancement in charge separation efficiency as quantified by the I-V and P-V characteristics indicates that coherence and excitonic delocalization might indeed be playing a crucial role in causing high efficiency of charge separation in the reaction center. 

\section{Emulation of the photosynthetic reaction center using a quantum dot based system}
\subsection{The idea}
Charge separation is a crucial step in photovoltaic devices~\citep{yu1995charge}. The mechanism involving excitonic delocalization for highly efficient charge separation in reaction center gives us pointers for the design of a highly efficient charge separation device. Out of all the solid state systems available, quantum dot seems to 
be viable candidate to realize such a device as GaN based quantum dots have been shown to host delocalized excitons~\citep{biolatti2000quantum,de2002intrinsic}.

A lot of progress has been made in quantum dot based realization of qubits and other quantum computing operations~\citep{loss1998quantum,de2002intrinsic,petta2005coherent}. Quantum dot based photovoltaic devices constitute an equally popular area of research~\citep{nozik2002quantum,aroutiounian2001quantum}. 
\begin{figure}[t]
\centering
\includegraphics[width = 6
cm,keepaspectratio]{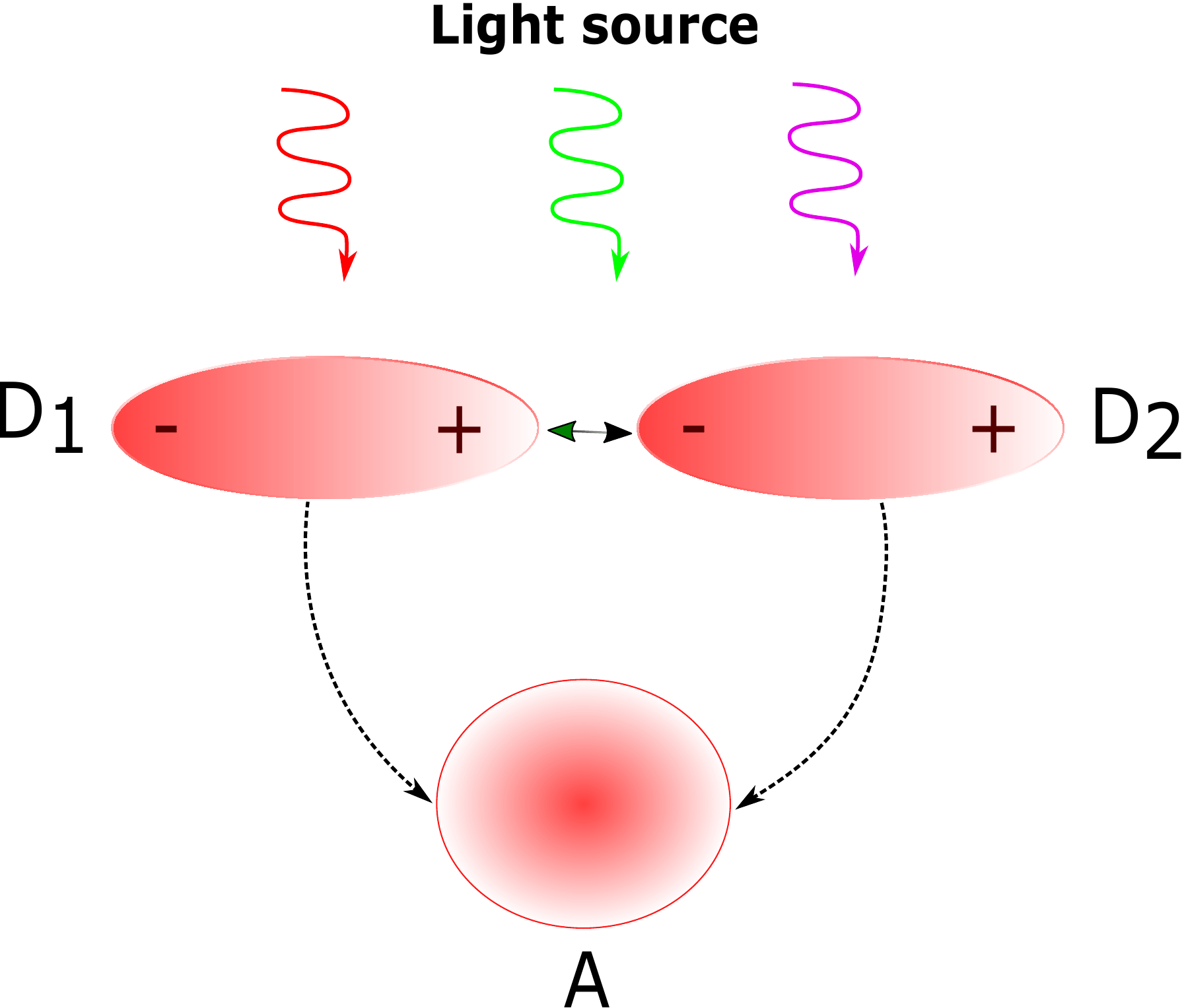}
\caption[A quantum dot based biomimetic system to emulate the charge separation mechanism in the photosynthetic reaction center]{A quantum dot based system that can potentially emulate the photosynthetic reaction center and enable high efficiency photon to charge conversion. It consists of two very closely spaced GaN quantum dots ($D_1$ and $D_2$). The in-built electric field delocalizes the excitons over them. They are coupled to a third quantum dot A via tunneling coupling. $D_1$ and $D_2$ transfers the electrons to quantum dot A via tunneling when there is difference in chemical potential across them. 
}
\label{QDs_RC}
\end{figure}
It turns out that these two areas of research can be bridged and the study of quantum effects in photosynthetic reaction center enables us to do this. The idea is as follows: quantum dots with built-in electric fields give rise to high excitonic dipole moment. This results in dipole-dipole coupling between the excitons if their host quantum dots are close enough. This coupling shifts the energy levels of the excitons and make them delocalized over the dots. Quantum dots based on GaN heterostructures exhibit built-in electric fields due to the spontaneous polarization and piezoelectric field which also makes them an ideal candidate to be used as qubits~\citep{de2002intrinsic}.  Most importantly, the dipole-dipole coupling between these quantum dots is of the same order as it is in the photosynthetic reaction center donor pair~\citep{de2002intrinsic}. Taking a cue from this, we propose a quantum dot based system where two closely spaced GaN quantum dots are illuminated with a light source resulting in generation of the excitons. The built-in electric field of the GaN quantum dots causes high dipole moment of their excitons and proximity of quantum dots enable strong dipole-dipole interaction between these dots. This delocalizes the excitons over both quantum dots; thus mimicking the special pair of the photosynthetic reaction center. These two quantum dots are then coupled to a nearby third quantum dot via tunnel-coupling that ensures electron transport from the donor pair to the acceptor~\citep{contreras2014dephasing}. This can be achieved by putting the system between appropriate source and drain contacts as has been proposed in Ref.~\citep{contreras2014dephasing}. This system looks very similar to the photosynthetic reaction center and can potentially emulate highly efficient charge separation mechanism of the reaction center. The schematic for this idea is shown in Fig.~\ref{QDs_RC}. The things that remain to be studied are: the formation of dark states after dipole-dipole interaction as has been mentioned for photosynthetic reaction center in Ref.~\citep{creatore2013efficient}, rates of electron transfer from donor quantum dots to the acceptor quantum dot which will depend on the strength of tunnel-coupling between the quantum dots, and the upconversion of the energy of solar photons to match the energy levels of the GaN quantum dots. In order to address the last challenge, other quantum dot systems should also be explored for hosting delocalized excitons.


\begin{figure}[htbp]
\centering
\includegraphics[width=1.0\linewidth]{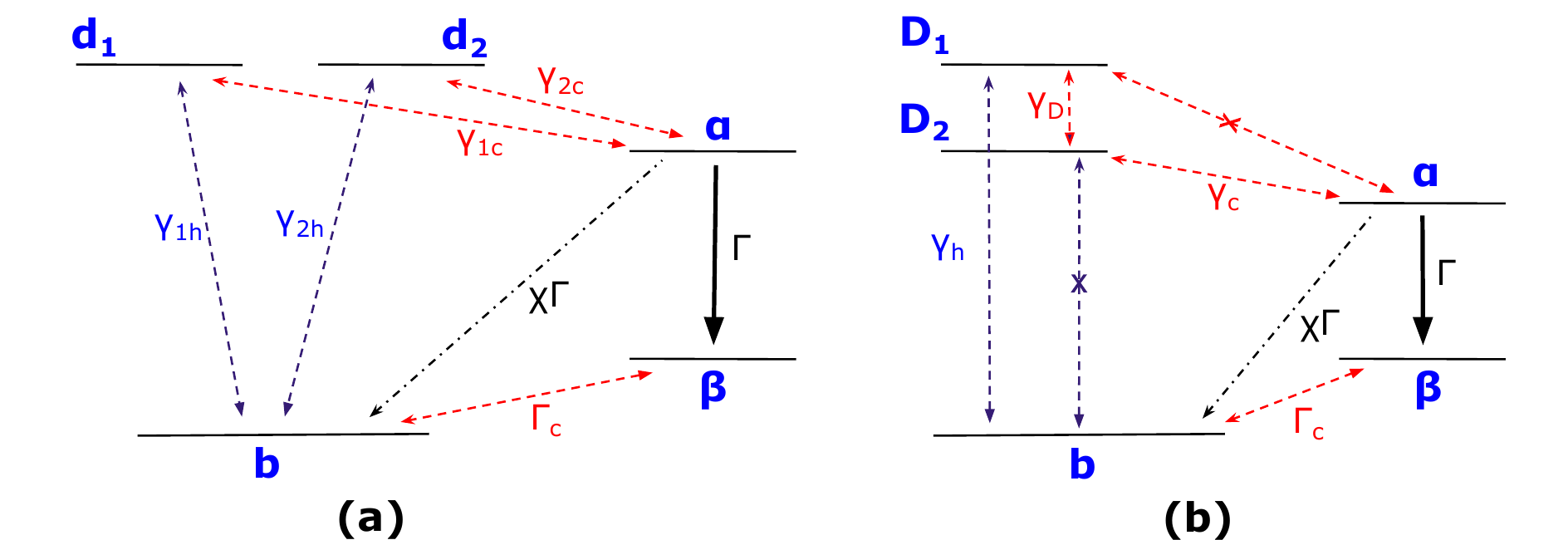}
\caption[State diagram of the quantum dot based biomimetic system]{State diagram of quantum dot system demonstrating its energy in various states.}
\label{QDs_EnergyLevels}
\end{figure}
\subsection{Calculations: effect of excitonic delocalization on charge separation efficiency of quantum dot system}
The state diagram for the quantum dot based system proposed to operate as coupled qubits~\citep{de2002intrinsic}, is shown in Fig.~\ref{QDs_EnergyLevels}. The energy levels for this system are:
$E_1 = 3.177$ eV, $E_2 = 3.255$ eV, and energy shift due to dipole-dipole interaction is: $\Delta E = -4.4$ meV. The values of the transition rates between energy levels of quantum dot systems is not available. In order to analyze the proposed system for photon to charge conversion operation, we take the transition rates from Ref.~\citep{creatore2013efficient} where a photosynthetic reaction center is studied.

\begin{figure}[htbp]
\centering
\includegraphics[width=5cm,keepaspectratio]{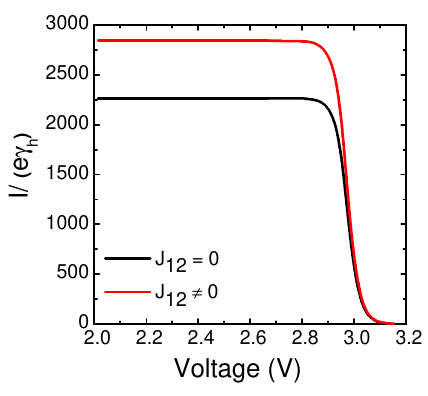}
\includegraphics[width=5cm,keepaspectratio]{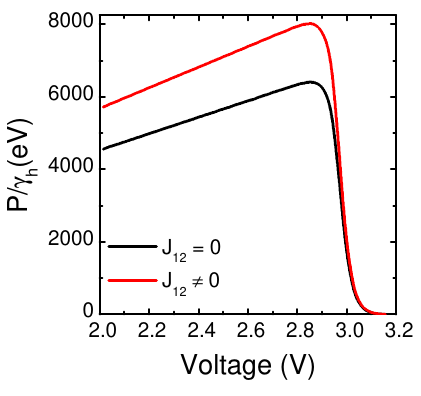}
\caption[I-V and P-V characteristics of the proposed system with and without coherence]{I-V and P-V characteristics of the proposed system with and without coherence. Presence of coherence results in 25.41\% improvement in the I-V characteristics and 25.18\% improvement in the P-V characteristics.}
\label{QDs_IV_PV}
\end{figure}
As in case of reaction center modeling, the role of excitonic delocalization on charge transfer is modeled by the I-V and P-V characteristics of the coupled quantum dots. The current, voltage and power are defined in exactly the same way as they are defined for the photosynthetic reaction center (cf. Eq.~\ref{RCQHE_Voltage} and Eq.~\ref{RCQHE_Current}).

The I-V and P-V characteristics of the system are given in Fig.~\ref{QDs_IV_PV} for both the cases when the excitons are localized (no dipole-dipole coupling) and delocalized (excitonic delocalization due to dipole-dipole coupling). It turns out that there is an improvement of 25.41\% in I-V characteristic of the system when excitons are delocalized as compared to the case when they are localized. Similarly, the P-V characteristics of the system also exhibit an improvement of 25.18\% when exciton are delocalized over the donor quantum dots as compared to the case when they are not.

\section{Discussion and Conclusion}
\label{Discussion}
Quantum coherence in the photosynthetic reaction center explains the high efficiency of charge separation in it. There have been several studies on exploring the ways in which coherence is sustained in this system and how it can be helpful in enhancing the charge separation efficiency. Two plausible ways in which coherence is sustained in the reaction center are: a) Fano interference, b) excitonic delocalization due to dipole-dipole coupling. Inspired by the dipole-dipole interaction induced efficiency enhancement in the reaction center, we have proposed a quantum dot based solid state system that can potentially emulate the photosynthetic reaction center and enable highly efficient photon to charge conversion. Although, transition dipole moments of excitonic states in GaN quantum dots after delocalization and electron transfer rate between neighboring quantum dots need to be studied before coming up with an actual device, the proposed scheme can potentially open up a new area of research for designing improved energy harvesting devices. In the future, our aim is to come up with a method of fabricating such devices. 
\acknowledgments{The author thanks Swaroop Ganguly for insightful comments about this work.}


\bibliography{RCEmulation}

\end{document}